\documentclass[twocolumn,showpacs,prl,aps]{revtex4}
\usepackage{graphicx}
\begin{document}
\title{Opinion dynamics: rise and fall of political parties}
\author{E.~Ben-Naim}\email{ebn@lanl.gov}
\affiliation{Theoretical Division and Center for Nonlinear Studies,
Los Alamos National Laboratory, Los Alamos, New Mexico 87545}
\begin{abstract}

We analyze the evolution of political organizations using a model in
which agents change their opinions via two competing mechanisms. Two
agents may interact and reach consensus, and additionally, individual
agents may spontaneously change their opinions by a random, diffusive
process. We find three distinct possibilities. For strong diffusion,
the distribution of opinions is uniform and no political organizations
(parties) are formed. For weak diffusion, parties do form and
furthermore, the political landscape continually evolves as small
parties merge into larger ones. Without diffusion, a pattern develops:
parties have the same size and they possess equal niches. These
phenomena are analyzed using pattern formation and scaling techniques.

\end{abstract}
\maketitle

Interacting particle systems and agent-based models are becoming
increasingly important in the behavioral, social, and political
sciences \cite{ww,tm}. There is compelling evidence that collective
phenomena emerging in social contexts can be attributed to basic
agent-agent interactions \cite{ra}. Statistical physics and nonlinear
dynamics methods naturally apply for analysis of simplified
interacting particle systems. Paradigms such as scaling, criticality,
and universality, emerging from such quantitative analysis, can guide
and validate detailed agent-based models, used to simulate real-world
situations.

In opinion dynamics, recent studies focus on the emergence of
cooperative phenomena including spatial organization, the formation of
coherent structures (political parties), and the transition from unity
to discord \cite{bfk,sg,szn,ds,sh,mr}. In particular, the remarkably
simple compromise process, in which pairs of agents reach a fair
compromise, captures familiar political systems: one-party, two-party,
etc, \cite{dnaw,wdan,hk,sm,sf,bkr,bkrv}.

This investigation generalizes the compromise process by allowing
individual agents to change their opinions in a random, diffusive
fashion. It is shown that diffusion is an essential element of opinion
dynamics. It generates realistic lifecycles of political organizations
and additionally, it governs the transition from a disorganized to an
organized political system.

The competition between compromise and diffusion is quantified by one
parameter, the diffusion constant. For strong randomness, the
political system is disorganized and the distribution of opinions is
uniform. For weak randomness, political organizations do form and they
evolve constantly. Large parties overtake smaller ones and the
separation between neighboring parties grows indefinitely. Without
randomness, a stationary pattern with evenly-spaced, evenly-sized
parties forms.

In our model, the opinion of an agent is quantified by an integer $n$,
and it changes via two separate processes. The first is
compromise. Two randomly selected agents reach consensus, provided
that their opinion difference is smaller than a fixed threshold, set
to two for simplicity,
\begin{equation}
\label{compromise}
(n-1,n+1)\buildrel 1\over \longrightarrow (n,n).
\end{equation}
The compromise process occurs at a constant rate, set to 1. The second
process is diffusion.  An agent may change his or her opinion in a
random fashion,
\begin{equation}
\label{diffusion}
n\buildrel D\over \longrightarrow n\pm 1.
\end{equation}
This is merely diffusion, a random walk in opinion space with $D$ the
diffusion constant. Of course, the total population is conserved. The
total opinion is strictly conserved in compromise events, but it is
conserved only on average for diffusive moves.

The compromise process mimics the human tendency for resolving
conflicts \cite{ra}, and the threshold incorporates a certain degree
of conviction in one's own opinion. Diffusion accounts for the
possibility that people may change their opinion either on their own
or due to news events, editorials, etc.

The density $P_n(t)$ of agents with opinion $n$ at time $t$ obeys the
master equation
\begin{eqnarray}
\label{master}
\frac{d P_n}{dt}
&=&2P_{n-1}P_{n+1}-P_n(P_{n-2}+P_{n+2})\\
&+&D(P_{n-1}+P_{n+1}-2P_n).\nonumber
\end{eqnarray}
The total population and the total opinion are conserved: $\sum_n
P_n={\rm const}$ and $\sum_n nP_n={\rm const}$. We analyze in order
one-party, two-party, and multi-party dynamics.

\noindent{\it One-party dynamics.} Consider the initial condition
$P_n(0)=m(\delta_{n,-1}+\delta_{n,0})$ with a well-defined political
organization, namely, a party.  Its size, equal to the total
population, is taken to be large, $m\gg D$. Clearly, throughout the
evolution, the opinion distribution remains symmetric,
$P_n=P_{1-n}$. Equation (\ref{master}) supports the trivial
steady-state where the opinion distribution vanishes, $P_n=0$, as well
as one in which compromise and diffusion balance
\begin{equation}
\label{steady}
P_{n-1}P_{n+1}=DP_n.
\end{equation}
Solving this equation recursively with $P_{-1}=P_0$ gives the periodic
state
\hbox{$(P_{-1},P_0,P_1,P_2,P_3,P_4)=(P_0,P_0,D,D^2/P_0,D^2/P_0,D)$},
with $P_n=P_{n+6}$. Starting with the one-party initial condition, the
distribution has a localized core that matches this periodic structure
over a few lattice sites,
\begin{eqnarray}
\label{core}
(P_0,P_1,P_2)\cong (m,D,D^2m^{-1}).
\end{eqnarray}
This is confirmed by numerical integration of (\ref{master}), as shown
in figure 1 \cite{onepeak}.  Using the conservation law
$P_0+P_1+P_2\cong m$, Eq.~(\ref{core}) can be refined
$(P_0,P_1,P_2)\cong (m-D,D,D^2m^{-1})$. The core is established very
quickly: from the short time behavior, $P_n(t)\cong m\,(Dt)^n$, we
immediately deduce the stabilization time scale $m^{-1}$. The larger
the party, the faster it is shaped. 

\begin{figure}[t]
\includegraphics*[width=0.45\textwidth]{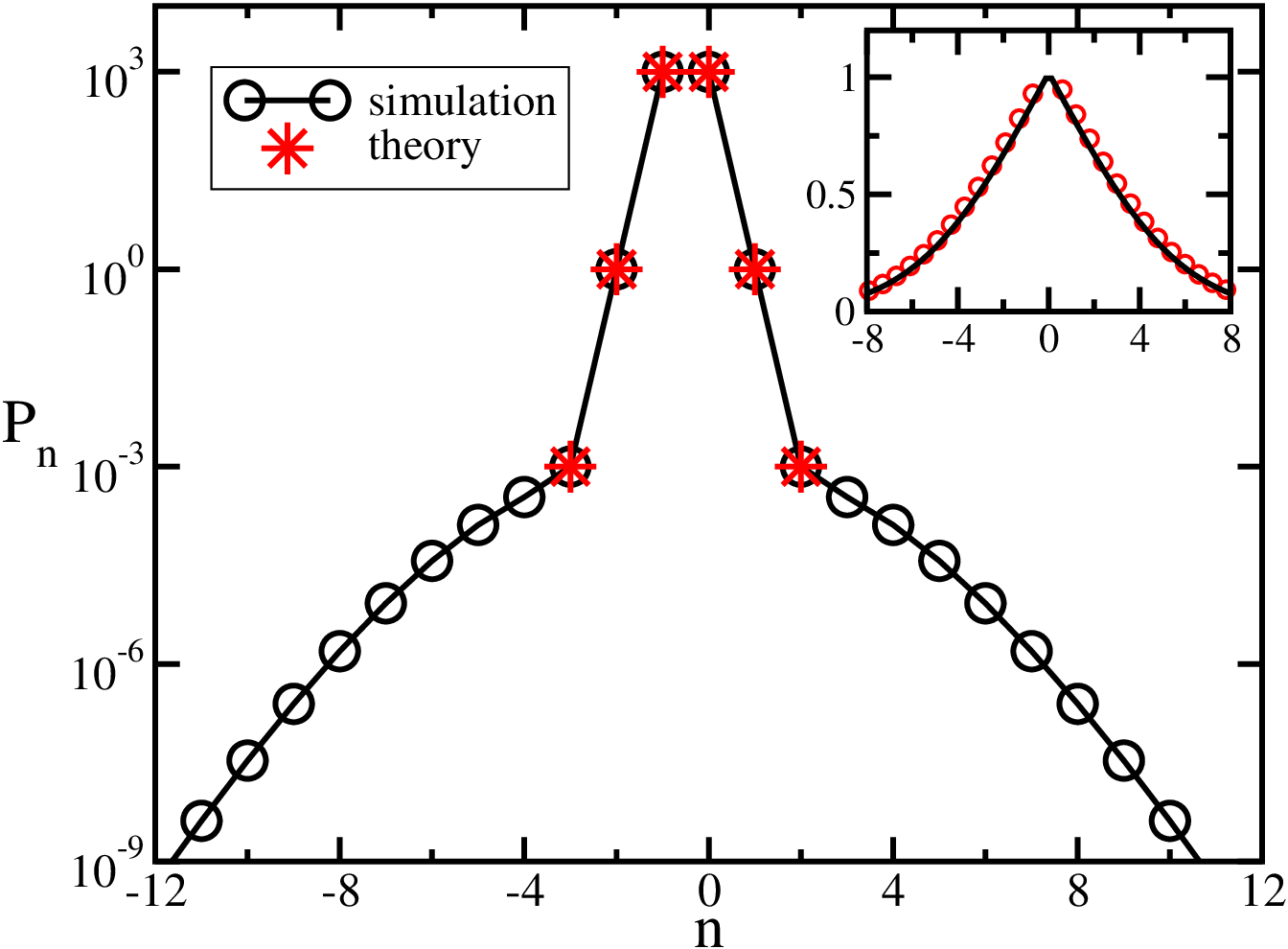}
\caption{One-party dynamics. Shown is $P_n(t=1)$ for $m=10^3$. The
theoretical prediction for the core (\ref{core}) is shown for
reference. The inset, where $\Phi(z)$ is plotted versus $z=nt^{-1/2}$,
shows the scaling behavior of the tail at large times $t=10^3$
(circles) and $t=10^4$ (line).}
\end{figure}

Outside the core, diffusion dominates over compromise since $P^2\ll
DP$ when $P\ll D$. The tail obeys the diffusion equation
\hbox{$dP_n/dt=D(P_{n-1}+P_{n+1}-2P_n)$} for $n\ge 2$ with the
boundary condition $P_2(t)=D^2m^{-1}$; this is the standard problem of
diffusion with a source \cite{cj}.  Consequently, the tail is
characterized by the diffusive length scale $\ell \sim t^{1/2}$ and
its shape is asymptotically self-similar
\begin{equation}
\label{tail}
P_n(t)\to m^{-1}\Phi\left(n\,t^{-1/2}\right)
\end{equation}
with $\Phi(0)={\rm const}$. Henceforth, the explicit dependence on the
diffusion constant is dropped. The tail population, $\mu=2\sum_{n\geq
2} P_n$, grows with time according to $\mu\sim m^{-1}t^{1/2}$.
Eventually, the entire population is transferred from the core to the
tail and the party dissolves. The lifetime of the party $\tau$ is
estimated from $\mu\sim m$ to be
\begin{equation}
\label{lifetime}
\tau\sim m^4.
\end{equation}
This time scale grows rapidly with the party size, indicating that
diffusive loss is negligible over a substantial period and that large
parties are long-lived. 

In summary, a single party has a quasi-stationary state consisting of
a fixed, tightly confined core and an extended diffusive tail.  Over
its lifetime, the core of the party is immobile and it is unaffected
by the random changes in position of its affiliates.  The core
contains the bulk of the population, its hight equals the party size
and its depth is inversely proportional to the size. Ultimately, an
isolated party dissolves. Its remnant is a diffusive cloud centered at
the original party position with total population equal to the initial
party size.

\noindent{\it Two-party dynamics.} To find out how two neighboring
parties interact, we consider the initial condition
$P_n=m_1(\delta_{n,-2}+\delta_{n,-3})+m_2(\delta_{n+2,l}+\delta_{n+3,l+1})$
corresponding to two large parties, $m_1,m_2\gg D$, that are separated
by a large distance $l\gg 1$.

\begin{figure}[t]
\includegraphics*[width=0.45\textwidth]{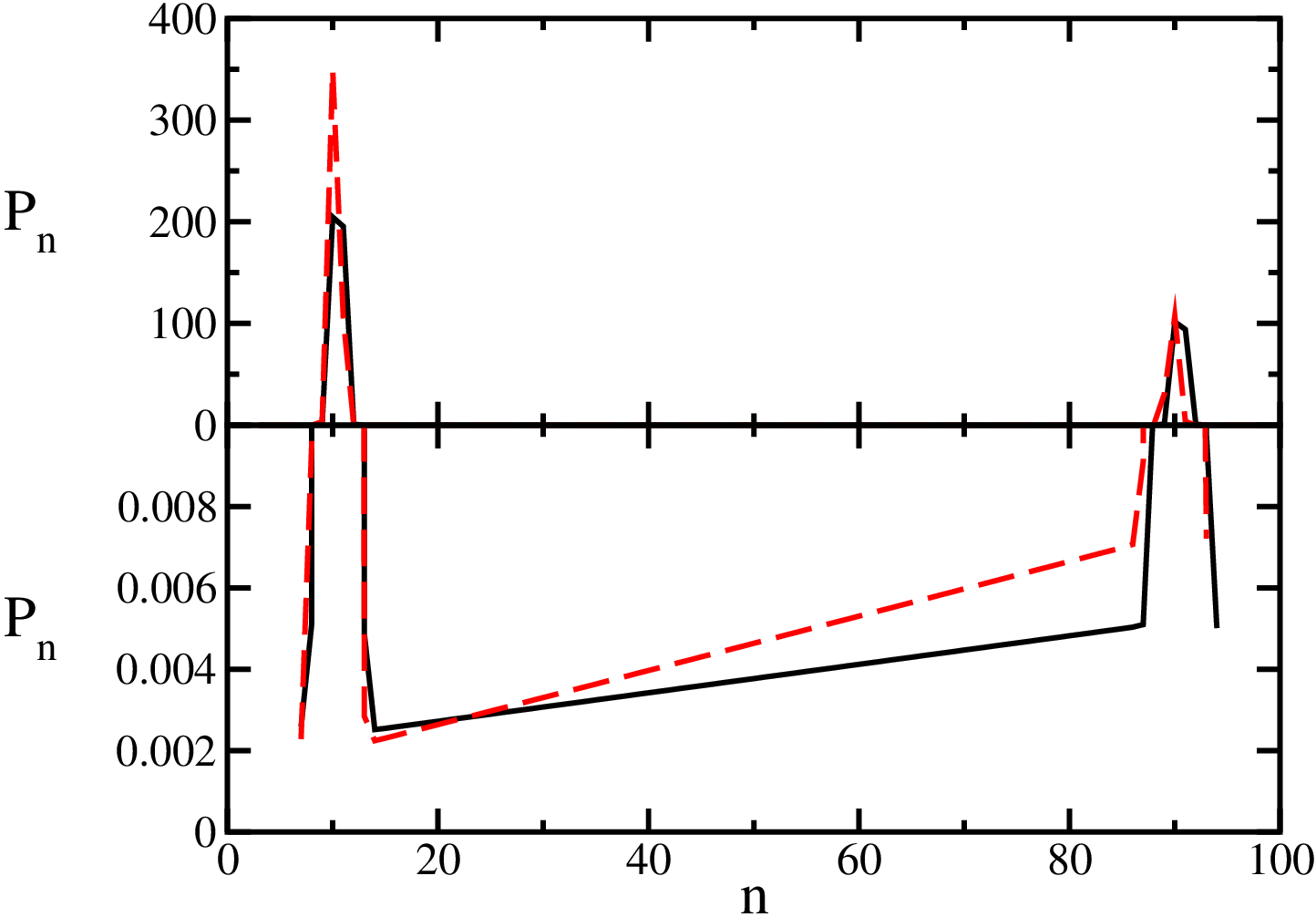}
\caption{Two-party dynamics.  Shown are results of numerical
integration with $m_1=2m_2=200$ and $l=90$ at an early time $t=10^3$
(solid lines) and a later time $t=2\times 10^5$ (dashed lines). The
top plot shows the cores, and the bottom plot shows the tail.}
\end{figure}

Initially, the parties do not interact and each one follows the
one-party dynamics above. When their diffusive tails meet, which
occurs on the diffusive time scale $l^2$, the distribution reaches a
steady-state in the region separating the two.  It obeys the discrete
Laplace equation \hbox{$P_{n+1}+P_{n-1}-2P_n=0$} with the boundary
conditions dictated by the two cores $P_0\propto m_1^{-1}$ and
$P_l\propto m_2^{-1}$. Therefore, there is a linear profile (figure 2)
\begin{equation}
\label{linear}
P_n\propto \frac{1}{m_1}+\left(\frac{1}{m_2}-\frac{1}{m_1}\right)\frac{n}{l}
\end{equation}
for $0\leq n\leq l$. As a result, there is a slow and steady flux from
the smaller party into the larger one, $J=|P_{n+1}-P_{n}|$, or
explicitly \hbox{$J\propto l^{-1}(m_<^{-1}-m_>^{-1})$} with $m_<={\rm
min}\,(m_1,m_2)$ and $m_>={\rm max}\,(m_1,m_2)$. We note that
diffusion enables the two parties to interact. The flux is
proportional to the difference in depth, and it is inversely
proportional to the separation. Eventually, the small party is
depleted. The depletion time can be estimated from the flux, $T\sim
m_</J$, as
\begin{equation}
\label{merger}
T\sim l\,m_<^2
\end{equation}
where the dependence on the larger population was tacitly ignored. An
improved estimate for the depletion time can be obtained from the
evolution equations \hbox{$dm_>/dt=-dm_</dt=J$}.

We conclude that there is a steady flux from the small party into its
neighboring larger party resulting in the eventual demise of the
smaller party. Thus, the result of the interaction is deterministic as
it always leads to merger. The lifetime of the small party grows
quadratically with its size. It also increases linearly with the
separation or ``niche''.  While the larger party grows during the
merger, this growth does not affect its position; it is practically
immobile.

\begin{figure}[t]
\includegraphics*[width=0.45\textwidth]{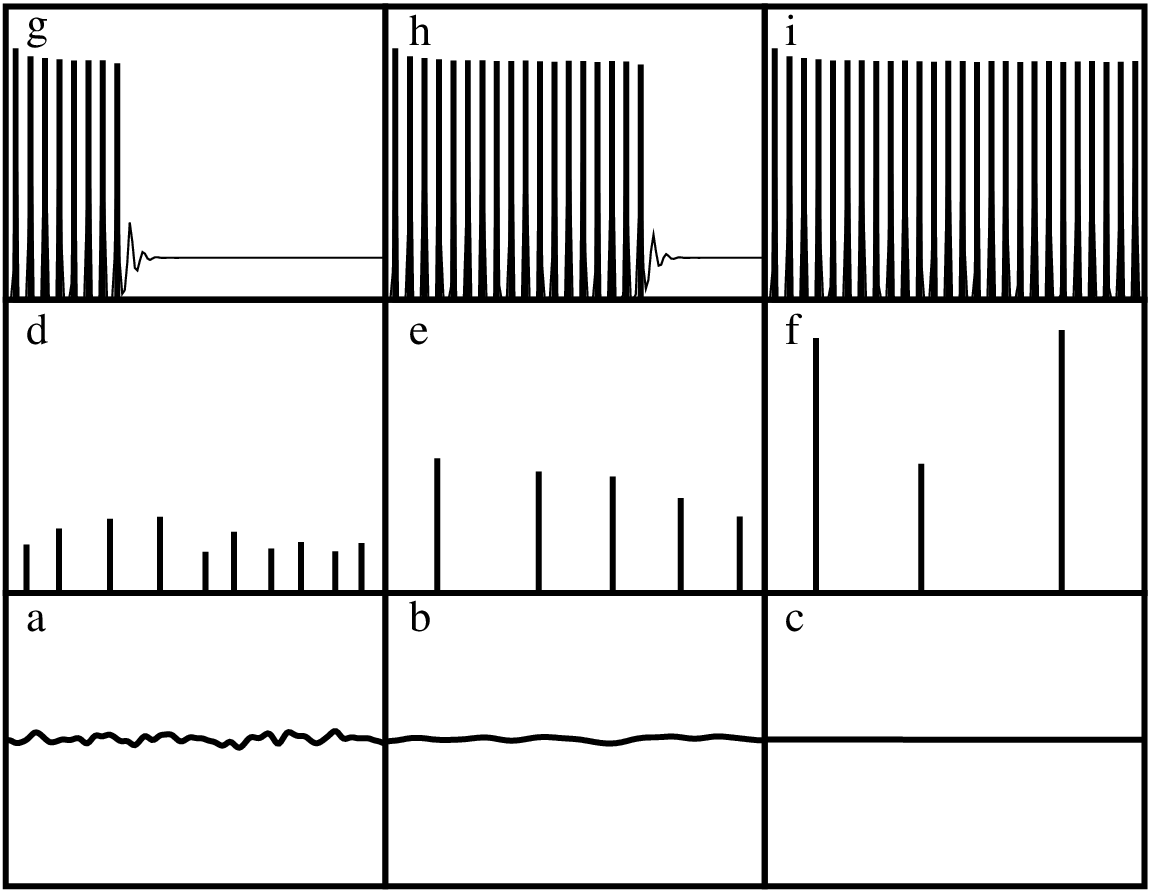}
\caption{Multi-party dynamics. Shown are representative snapshots for
$D=3$ (a-c), $D=1$ (d-f), and $D=0$ (g-i) at an early time (left),
intermediate time (middle), and late time (right). Shown is $P_n$
versus $n$ using lines ($D=3$, $D=0$). Also shown is the total party
size and position using bars with hight equal to the party size
($D=1$, $D=0$). The plots are results of numerical integration of
(\ref{master}) with $\epsilon=0.1$ for $D\ne 0$ and $\epsilon=0$ for
$D=0$.}
\end{figure}

\noindent{\it Multi-party dynamics.} The uniform state $P_n={\rm
const}$ is stationary. Any uniform state can be transformed by an
appropriate rescaling of the opinion distribution, time, and the
diffusion constant into the state $P_n=1$, so we address this case. To
investigate the stability of the uniform state, we considered
heterogeneous initial conditions with $P_n(0)$ a randomly chosen
number in the range $[-1-\epsilon,1+\epsilon]$ with $\epsilon\ll
1$. The system is large, $1\leq n\le N$ with $N\gg 1$.

Stability of the uniform state is studied using small periodic
perturbations, $P_n=1+\phi_n$, with $\phi_n\propto e^{ikn+\lambda t}$.
From (\ref{master}), the growth rate is
\begin{equation}
\label{growth}
\lambda=2(2\cos k -\cos 2k-1)+2D(\cos k-1).
\end{equation}
The perturbation decays if its wave-number is sufficiently large,
$|k|>k_0$ with $k_0=\cos^{-1}(D/2)$. 

\noindent{\it Uniform Distribution.} Since $k_0=0$ for $D=2$, there
is a critical diffusivity
\begin{equation}
\label{critical}
D_c=2.
\end{equation}
For strong diffusion, $D>D_c$, perturbations decay exponentially with
time and the uniform state is rapidly restored, regardless of the
initial conditions (figures 3a-3c).  Just above the critical
diffusivity, long wavelength perturbations are long lived: their decay
time diverges $\propto (D-D_c)^{-2}$, following from $\lambda \propto
(D-D_c)\,k^{2}$ as $D\downarrow D_c$. In any case, compromise
interactions become irrelevant and diffusion dominates. As a result,
the opinion distribution approaches a structureless state: the
political system is disorganized as no parties are formed.

\begin{figure}[t]
\includegraphics*[width=0.45\textwidth]{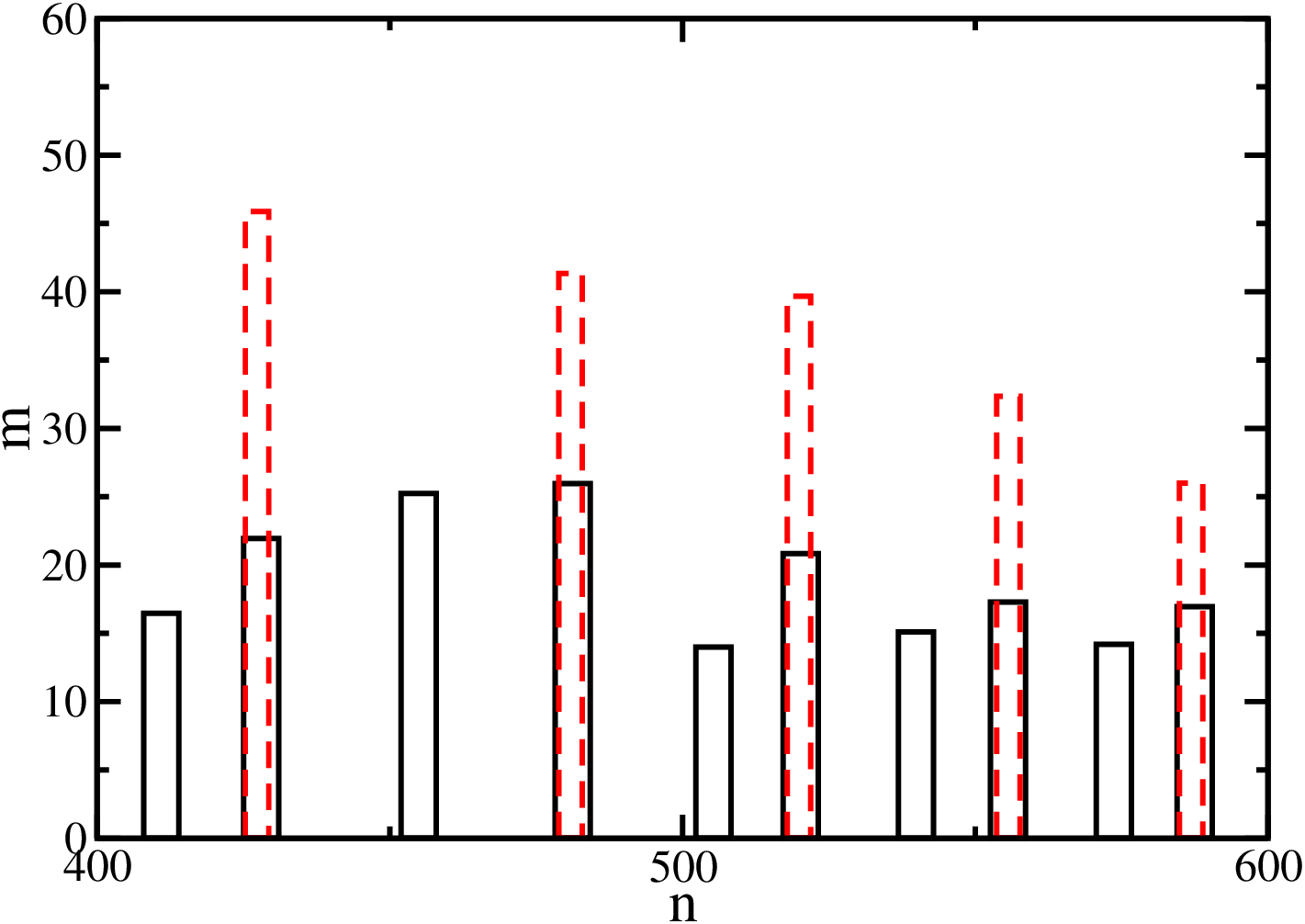}
\caption{The party size $m$ versus position $n$. Shown are results of
numerical integration of (\ref{master}) with $D=1$, $N=10^3$, and
$\epsilon=0.1$ at times $t=10^3$ (solid line) and $t=10^4$ (dashed
line). The bar hight equals the party size.}
\end{figure}

\noindent{\it Coarsening.} For weak diffusion, $D<D_c$, perturbations
to the initial state are magnified and parties are quickly formed
(figure 3d-3f). The system develops a mosaic of parties. Since the
initial state is heterogeneous, the size of the parties and the
separation between them vary.  The evolution follows straightforwardly
from the two-party dynamics and there is a linear profile in regions
separating parties. Small parties merge into larger neighboring
parties, and as a result, the remaining parties grow in size and in
niche (figure 4).

Let us assume that the typical size is $m$.  The population density
must be constant, so the typical niche is of the same order, $l\sim
m$. Substituting these scales into the depletion time (\ref{merger})
yields $T\sim m^3$ and since time is the only relevant time scale, the
typical size growth law is (figure 5)
\begin{equation}
\label{mass}
m\sim t^{1/3}.
\end{equation}
Asymptotically, the system reaches a self-similar state where the
party size is characterized by the typical scale $m$ and consequently,
the party size distribution, $Q_m$, becomes self-similar in the long
time limit, \hbox{$Q_m\sim t^{-1/3}\Psi(mt^{-1/3})$}. Our numerical
simulations confirm this, along with (\ref{mass}).

\begin{figure}[t]
\includegraphics*[width=0.45\textwidth]{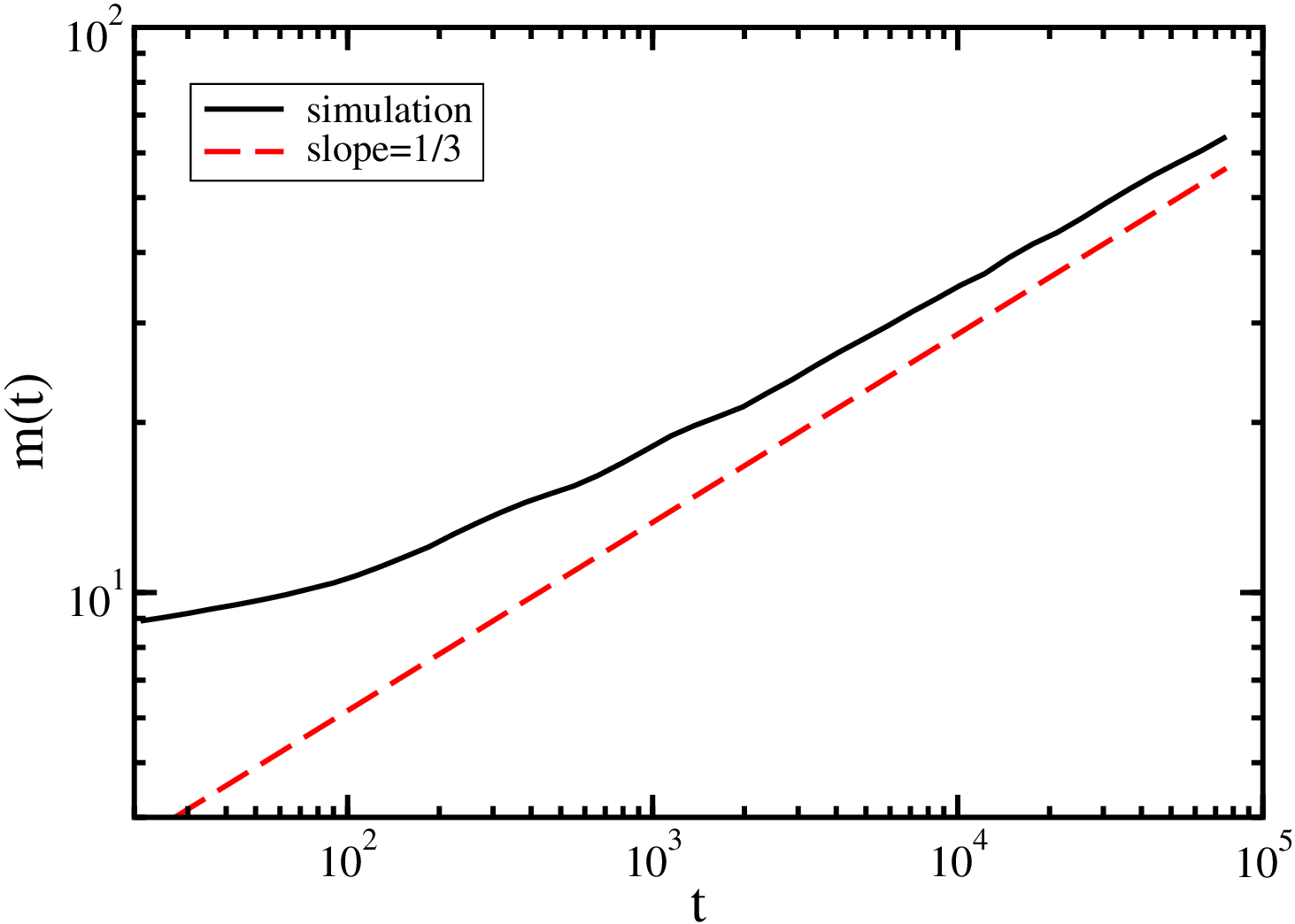}
\caption{The average party size versus time. Shown are simulation
results (solid line) obtained using $D=1$ and $N=10^5$ and a line of
slope $=1/3$ (dashed line) for reference.}
\end{figure}

The lifetime of a party is governed by the interplay between size and
niche. Typically, the larger the party, the longer it survives, but a
small party may still outlast a larger neighbor if it has a large
enough niche. Except for this size-niche competition, the coarsening
mechanism is similar to Lifshitz-Slyozov ripening \cite{ls}.

\noindent{\it Pattern Formation.} Without diffusion, the system
approaches a state where $P_{n-1}P_{n+1}=0$ for all $n$; there is no
evolution and parties are localized to either one or two sites
\cite{bkr}. For a narrow political spectrum, consensus is reached and
there is a single party. As the size of the spectrum increases, the
number of parties undergoes a series of bifurcations and there may be
two off-center parties, three parties, etc \cite{dnaw,bkr}.

Asymptotically, the number of parties grows linearly with the
political spectrum $N$ because the parties are equally-spaced (figure
3i). The spacing equals the party size, as follows from conservation
of population. This pattern formation can be understood using linear
stability analysis. We demonstrate this for uniform initial
distributions.

Consider the uniform initial condition: $P_n(0)=0$ for $n<0$ and
$P_n(0)=1$ for $n\geq 0$. This state is unstable with respect to
perturbations that propagate from the boundary into the unstable
uniform state (figures 3g-3i). A small periodic disturbance
$P_n=1+\phi_n$ with $\phi_n\propto \exp[i(kn-\omega t)]$ is
characterized by the dispersion relation $\omega=2i(2\cos k-\cos 2k
-1)$ according to the diffusion-free evolution equation
(\ref{master}). A saddle point analysis shows that the propagation
velocity $v$ obeys \cite{vvs}
\begin{equation}
\label{velocity}
v=\frac{d\omega}{dk}=\frac{{\rm Im}[w]}{{\rm Im}[k]}.
\end{equation}
The solution is $k=k_{\rm select}+i\lambda$ with the selected wave
number $k_{\rm select}=1.183032$. The decay constant $\lambda=0.467227$
characterizes the exponential decay far into the unstable state,
$\phi_n\sim \exp[-\lambda (n-vt)]$. The propagation velocity is
$v=3.807397$ and the period of the pattern is $L_{\rm
select}=2\pi/k_{\rm select}=5.311086$. Our numerical studies confirm
these results.

In the absence of boundaries, i.e., in a periodic system, the
perturbation with the largest growth rate dominates and sets the
wavelength.  From (\ref{growth}), the growth rate is $\lambda=2(2\cos
k -\cos 2k-1)$; it is periodic in $k$, $\lambda(k)=\lambda(k+2\pi)$.
The uniform state is unstable with respect to long wavelength
perturbations with $k<\pi/2$.  Also, the growth rate is maximal at
$k_{\rm max}=\pi/3$ and the corresponding period, $L_{\rm
max}=2\pi/k_{\rm max}$, is $L_{\rm max}=6$.

Numerically, we find that the actual period falls between the two
linear stability values
\begin{equation}
\label{period}
L\approx 5.67.
\end{equation}
Starting from a compact distribution, perturbations with the selected
period are generated by the boundary and they propagate into the
interior.  As the disturbance reaches the interior of the system,
perturbations with a smaller wavelength, that have a larger growth
rate, dominate.  This argument suggests the above estimates as bounds
for the period $L_{\rm select}<L<L_{\rm max}$.  These bounds are
tight, so linear stability analysis yields a very good approximation
for the period. Yet, the pattern selection mechanism is intrinsically
nonlinear and obtaining the exact period remains a challenge.

In summary, we found that the level of noise (diffusion) determines
the nature of the political system. Strong noise leads to a uniform
distribution of opinions with every possible opinion equal in
weight. With weak noise, the system organizes into political parties
and the political landscape undergoes coarsening with large parties
continuously overtaking small ones. Without noise, the system evolves
into a frozen pattern of parties with equal weights and equal
separations.

Several qualitative features are surprisingly realistic.  The
lifecycle of a party includes formation, growth, and demise. Isolated
parties have a fixed position and their lifetime grows rapidly with
their size, but ultimately, any party dissolves.  When two parties
interact, the smaller party loses ground to the larger party at a
steady rate. A small party with a large niche can be long lived.

Diffusion plays a critical role. It facilitates interaction between
parties and it is responsible for the dissolution of
parties. Moreover, the patterned state is unstable with respect to
addition of diffusion.  We conclude that spontaneous opinion changes
are an integral part of opinion dynamics.

I thank M.~Cross, P.~L.~Krapivksy, S.~Redner, and L.~Tsimring for
useful discussions. I also acknowledge DOE grant W-7405-ENG-36 for
support of this work.


\begin{thebibliography}{99}

\bibitem{ww}
   W.~Weidlich, {\it Sociodynamics: A Systematic Approach to Mathematical
   Modelling in the Social Sciences}
   (Harwood Academic Publishers, 2000).

\bibitem{tm}
   T.~M.~Liggett, {\it Interacting particle systems}
   (Springer-Verlag, New York, 1985).

\bibitem{ra}
  R.~Axelrod, {\it The complexity of cooperation} (Princeton University
  Press, 1997).

\bibitem{bfk}
      E.~Ben-Naim, L.~Frachebourg, and P.~L.~Krapivsky,
      Phys. Rev. E {\bf 53}, 3078 (1996).

\bibitem{sg}
   S.~Galam, 
   Physica A {\bf 238}, 66 (1997).

\bibitem{szn}
   K.~Sznajd-Weron and J.~Sznajd, Int.\ J. Mod.\ Phys.\ C {\bf 11},
   1157 (2000).

\bibitem{ds} 
     D.~Stauffer, J. Artif.\ Soc.\ Soc.\ Simul.\ {\bf 5}, 477
    (2002).

\bibitem{sh}
   A.~Soulier and T.~Halpin-Healy, 
   Phys. Rev. Lett. {\bf 90}, 258103 (2003).

\bibitem{mr}
   M.~Mobilia and S.~Redner,
   Phys. Rev. E {\bf 68}, 046106 (2003).

\bibitem{dnaw}
   G.~Deffuant, D.~Neau, F.~Amblard and G.~Weisbuch,
   Adv. Comp. Sys. {\bf 3}, 87 (2000).

\bibitem{wdan}
   G.~Weisbuch, G.~Deffuant, F.~Amblard, and J.~P.~Nadal,
   Complexity {\bf 7}, 55 (2002).

\bibitem{hk}
   R.~Hegselman and U.~Krause,
   J. Art. Soc. Soc. Sim. {\bf 5}, 2 (2002). 

\bibitem{sm}
    D.~Stauffer and H.~Meyer-Ortmanns,
    Int. J. Mod. Phys. B {\bf 15}, 241 (2004).

\bibitem{sf}
   S.~Fortunato,
   cond-mat/0406054. 

\bibitem{bkr}
      E.~Ben-Naim, P.~L.~Krapivsky, and S.~Redner,
      Physica D {\bf 183}, 190 (2003)

\bibitem{bkrv}
     E.~Ben-Naim, P.~L.~Krapivsky, R.~Vazquez, and S.~Redner,
     Physica A {\bf 330}, 99 (2003). 

\bibitem{onepeak} For $P_n(0)=m\delta_{n,0}$ the core has a different
    shape
    $(P_0,P_1,P_2,P_3)=(m,D^{1/2}m^{-1/2},D^{3/2}m^{1/2},D^2m^{-1})$, but 
    essential features (height and depth) are the same. 

\bibitem{cj}
    H.~S.~Carslaw and J.~C.~Jaeger,
    {\it Conduction of Heat in Solids}, 
    (Oxford University Press, New York 1959).

\bibitem{ls}
       I.~M.~Lifshitz and V.~V.~Slyozov,
       Zh.\ Eksp.\ Teor.\ Fiz. {\bf 35}, 479 (1959)
       [Sov.\ Phys.\ JETP {\bf 8}, 331 (1959)].

\bibitem{vvs}
   W.~van Saarloos, 
   Phys. Rep. {\bf 386}, 29 (2003).

\end{thebibliography}
\end{document}